\documentclass[letter,twocolumn,10pt]{article} 
\usepackage{usenix}

\usepackage{lipsum} 
\usepackage{amssymb}
\usepackage{amsmath}
\usepackage{verbatim}
\usepackage[hyphens]{url}
\usepackage[breaklinks]{hyperref}  
\usepackage{pslatex}
\usepackage{array}
\usepackage{subcaption}
\usepackage{pdflscape}
\usepackage{textcomp}
\usepackage{algorithmic}
\usepackage{graphicx}
\usepackage{wrapfig}
\usepackage{framed}
\usepackage{booktabs} 
\usepackage{amsthm} 
\usepackage{epigraph} 
\usepackage[textsize=small,shadow]{todonotes}  
\usepackage{listings} 
\usepackage{color}
\usepackage[ampersand]{easylist}  
\usepackage{anyfontsize}

\usepackage[citestyle= numeric-comp,
			bibstyle=numeric,
            sorting=none,
	        natbib=true,
            hyperref=true,
            doi=false,
            url=true,
            isbn=false,
            eprint=false,
            maxcitenames=2,
            backend=bibtex]
            {biblatex}

\usepackage{etoolbox}
\AtEveryBibitem{%
  \ifentrytype{misc}
  {
  }
  {
    \clearfield{url}%
    \clearfield{urldate}%
  }%
}
\bibliography{./library}

\hypersetup{
	colorlinks=true,
	citecolor=red,
	linkcolor=blue,
	urlcolor=blue
}
\begin{document}
\newtheorem*{assumption}{Assumption}
\newtheorem*{qnn}{Research Question}
\newtheorem{q}{Research Question}
\newtheorem{subq}{Sub-Question}[q]
\newtheorem*{guessnn}{Hypothesis}
\newtheorem{guess}{Hypothesis}
\newtheoremstyle{mystyle}
  {}
  {}
  {\itshape}
  {}
  {\normalfont}
  {.}
  { }
  {}
\theoremstyle{mystyle}
\newtheorem{subguess}{\textit{\underline{Hypothesis}}}[subq]
\newtheorem*{expl}{\textit{Explanation:}}

\newcounter{mycomment}
\newcommand{\mycomment}[2][]
           {\refstepcounter{mycomment}
           {\todo[inline,color={red!100!green!33},size=\small,caption={}]
           {\textbf{Comment [\uppercase{#1}\themycomment]:}~#2}}}
\newcounter{mytodo}
\newcommand{\mytodo}[2][]
           {\refstepcounter{mytodo}
           {\todo[inline,size=\small]
           {\textbf{TODO [\uppercase{#1}\themytodo]:}~#2}}}

\title{Developing Security Reputation Metrics for Hosting Providers\\\normalsize (Step Right Up! Metrics Galore! )}


\urldef{\mails}\path|{A.Noroozian,Maciej.Korczynski,S.T.Tajalizadehkhoob,M.J.G.vanEeten }@tudelft.nl|
\author{
{\rm A. Noroozian}\\
\and
{\rm M. Korczy\'nski}\\
\and
{\rm S. TajalizadehKhoob}\\
\and
{\rm M. van Eeten}\\
\and
{\rm Delft University of Technology}\\}

\index{Noroozian, Arman}
\index{Korczy\'nski, Maciej}
\index{TajalizadehKhoob, Samaneh}
\index{van Eeten, Michel}


\maketitle


%
%

\begin{abstract}
Research into cybercrime often points to concentrations of abuse at certain hosting providers. 
The implication is that these providers are worse in terms of security; some are considered `bad' or even `bullet proof'.

Remarkably little work exists on systematically comparing the security performance of providers. 
Existing metrics typically count instances of abuse and sometimes normalize these counts by taking into account the advertised address space of the provider. 
None of these attempts have worked through the serious methodological challenges that plague metric design. 

In this paper we present a systematic approach for metrics development and identify the main challenges: (\textbf{\textit{i}})~identification of providers, (\textbf{\textit{ii}})~abuse data coverage and quality, (\textbf{\textit{iii}})~normalization, (\textbf{\textit{iv}})~aggregation and (\textbf{\textit{v}})~metric interpretation. 
We describe a pragmatic approach to deal with these challenges. 
In the process, we answer an urgent question posed to us by the Dutch police: `which are the worst providers in our jurisdiction?'.
Notwithstanding their limitations, there is a clear need for security metrics for hosting providers in the fight against cybercrime.
\end{abstract}
%
%
\pagestyle{plain}
\section{Introduction}\label{sec:introduction}
\vspace{-7pt}
Hosting providers are companies that provide servers via which customers can make content or services available on the Internet \textit{e.g.} websites, email or support for multi-player gaming. 
As with virtually all services on the Internet, they are abused for criminal purposes as well. 
A wealth of research has identified how hosting infrastructure shows up in various criminal business models. 
Think of phishing sites, command-and-control servers for botnets, child pornography, malware distribution, and spam servers~\cite{Stone-Gross2009}.

Nobody contests that hosting providers play a key role in fighting cybercrime. 
Much of the criminal activity runs on compromised servers of legitimate customers, some on servers rented by the criminals themselves. 
In either case, the hosting providers typically becomes aware of the problem only after being notified of the abuse. 
Their response to abuse reports varies widely, ranging from vigilant to slow to negligent or even bullet-proof~\cite{Canali2013b,Stone-Gross2009}. 
To empirically measure which of these responses is actually occurring has proven to be very challenging. 
Existing metrics of hosting provider security typically count instances of abuse within an Autonomous System, sometimes normalized by the size of the advertised address space~\cite{Stone-Gross2009, Shue2012,hostexploit}. 
None of these attempts adequately account for the serious methodological challenges plaguing such metrics.

In this paper, we present a systematic approach for developing metrics for hosting providers. 
It enables us to identify and discuss the main challenges: (\textbf{i\textit{}})~identification of providers, (\textbf{\textit{ii}})~abuse data coverage and quality, (\textbf{iii\textit{}})~normalization, (\textbf{iv\textit{}})~aggregation and (\textbf{v\textit{}})~metric interpretation in light of the heterogeneity of hosting providers. 
Additionally we present a pragmatic approach to deal with these issues. 

This study is part of an ongoing collaboration with the Dutch National High Tech Crime Police, the Authority for Consumers and Markets, the Public Prosecutor and the Dutch Hosting Provider Association. 
The objective is to answer an urgent question posed by the police: which are the worst providers in our jurisdiction? 

The question illustrates there is a clear need for security metrics for hosting providers, notwithstanding their limitations. 
Reducing cybercrime is as much a problem of incentives as it is a technical issue~\cite{Moore2009a}. 
Without reliable signals we cannot tell which provider is vigilant, lax, negligent or outright criminal and it will be very difficult to move the sector towards more secure practices. 
\textit{Information asymmetry} erodes the incentives of providers to invest in security. 
Reliable metrics can (i) signal security performance to customers, upstream and downstream providers, law enforcement and other stakeholders (ii) enable benchmarking of providers, and (iii) help identify the effectiveness of security practices and policies.

The main contributions of this paper are as follows: (i) we systematically outline the process to develop security reputation metrics for hosting providers, as well as the methodological challenges encountered along the way, (ii) we improve existing techniques for mapping abuse to hosting providers and for taking into account the size of hosting providers in computing reputation scores, (iii) we present a pragmatic approach to produce metrics for the Dutch market, developed in collaboration with key stakeholders.
\section{Background}
\vspace{-7pt}
\begin{figure*}[thb]
\includegraphics[width=1\linewidth]{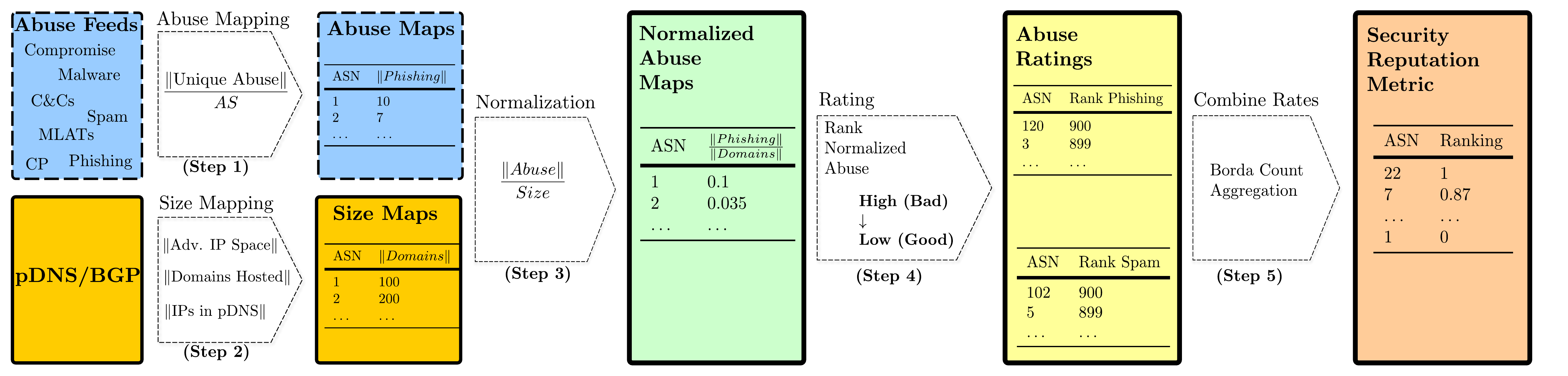}
\caption{Reputation metric development process}
\label{fig:process_all}
\vspace{-10pt}
\end{figure*}

Hosting providers come in many shapes and sizes and offer portfolios of services: from relatively expensive dedicated physical machines to virtual private servers (VPS) to the cheaper options of shared hosting or even so-called free hosting.
In each service, the role of the provider vis a vis the customer is different. On a dedicated machine, and to a lesser extent on a VPS, the customer controls the entire software stack, whereas on shared hosting, many customers operate under restricted privileges on a machine they share with many other users. 
Free hosting services limit user control to the extreme. 

Depending on the type of customer, hosting providers play a different role in protecting their customers against compromise by patching servers, cleaning and monitoring for abuse. 
Similarly, providers need to protect the rest of the Internet against potentially malicious customers by putting in place different checks and restrictions which depends on the service contract with that customer.

Next to the occurrence rate of abuse, the uptime of abuse also reflects hosting provider security practices. 
On one end of the spectrum, vigilant hosting providers remove malicious content often within hours of its discovery, in the middle there are some providers that respond more slowly and more selectively and on the other extreme are the so called `bulletproof' hosting providers that seem to ignore all abuse notifications. 

There has been a lot of speculation over the incentives of providers. 
A shared hosting provider, for example, could act against abuse more directly because its customers have only limited control over the machines that they use. 
On the other hand, shared hosting is a highly competitive market with low margins, so investing in security is not likely to be a high priority. 
The only way forward is to replace speculation with reliable empirical evidence of abuse rates across providers.
\section{Overview of Approach}\label{sec:overview}
\vspace{-7pt}
Our approach for calculating reputation metrics is partly guided by the goal to allow our collaborators to engage in meaningful discussions based on reliable empirical techniques. 
To this end, we produce two types of security indicators for hosting providers based on data available in public and private abuse feeds: (\textbf{\textit{i}})~\textit{Occurrence of abuse}: an indicator based on counting occurrence of abuse, and (\textbf{\textit{ii}})~\textit{Persistence of abuse}: an indicator based on how long the abuse was present. 
Distinguishing between occurrence of abuse and the response of a hosting provider to abuse as independent measures of performance is important. 
While the occurrence of abuse is to some extent inevitable due to technical vulnerabilities and related to organization size and attacker characteristics, persistence of abuse indicates attitude towards dealing with abuse and mainly relates to defender characteristics. 
In conjunction, these independent indicators provide a better understanding of the overall security performance of a hosting provider. 

Figure~\ref{fig:process_all} illustrates a high level overview of the complete procedure to produce these indicators. 
Here, boxes represent inputs/outputs to each step while arrows transformation steps on data required for each step. 
The process is quite generic and outlines the steps that any reputation metric requires to arrive at final scores. 
In executing these steps there are challenges that need to be overcome and choices that have to be made that will undoubtedly effect the reliability and interpretation of the reputation metric.
 In what follows we systematically walk the reader through the steps of the process meanwhile highlighting challenges related to each step and the possible effects on the overall metric and its interpretation. 
A more detailed analysis of some choices and their effect are presented later in Section~\ref{sec:sensitivity}.
\section{Step 1 -  Abuse Mapping}\label{sec:abuse_mapping}
\vspace{-7pt}
Identifying hosting providers is not straight forward since they do not directly map onto entities with which underlying Internet protocols work or what abuse data capture.
The first decision that needs to be made is to \textit{identify what a hosting provider is}. 
\vspace{-13pt}

\subsection{Identifying Hosting Providers}
\vspace{-3pt}
To produce reputation metrics for Dutch hosting providers, we made the (common~\cite{Stone-Gross2009,Shue2012,Wagner2013,Bradbury2014}) assumption that hosting providers will have an associated Autonomous System Number (ASN). 
Consequently, we considered any AS which routed IP addresses that geo-located to the Netherlands as a Dutch hosting provider.

While the assumption may hold in general, ASes could refer to Internet Service Providers (ISP), Internet exchange points, banks, governmental institutions, universities and in general non-hosting entities. 
Without a deeper analysis of the ASes, such an assumption may lead to considerable error in mapping abuse onto hosting.   
Even when an AS does refer to a hosting provider, complexity still exists.  
Providers may have multiple ASNs or there might be multiple organizations which own a part of the IP space in an AS or \textit{resellers} who lease infrastructure from the AS.
Some ASes also route traffic destined for IPs owned by \textit{peers}.
Certain legitimate services (\textit{e.g.} CloudFlare) may act as proxies and hide providers. 
As a result abuse associated with small organizations with registered IPs in ASes may end up attributed to the AS from which the infrastructure is leased. 
We have opted to analyze abuse for organizations with registered IPs inside ASes in future work.

One method to better identify hosting providers and identifying organizations under each AS, is to analyze IP `\textit{ownership}' using Maxmind's GeoIP ISP Database.  
Utilizing such information results in a more fine grained mapping which mitigates the mapping problems discussed above. 
Nevertheless, this approach has complications of its own such as non-standardized WHOIS data formats where the same organization might appear with multiple names that are non-trivial to relate to each other.
For example, the Dutch provider \verb|Leaseweb| might appear under any of the following additional names:   \verb|Leaseweb Asia Pacific. ltd.|, \verb|leaseweb1.iomadserve.com|.
\vspace{-20pt}
\subsection{Unit of Abuse}
\vspace{-15pt}
The second key decision is about the \textit{unit of abuse} or \textit{how to count the abuse data}.
Unlike other hosting metrics which typically count distinct IP addresses as the unit for abuse~\cite{Stone-Gross2009,Shue2012,Bradbury2014}, our approach considers unique 2nd-level domain-IP pairs -  $\left\langle  2LD, IP \right\rangle$ -  as the unit of abuse. 
From this point on, we use the terms `2LD' and `domain' interchangeably, unless the context requires otherwise.
Simply counting the number of abusive IP addresses largely underestimates abuse from shared hosting services since criminals may use the same IPs for various purposes.
For example a compromised server may host a phishing website and also be used for spreading malware. 
Furthermore, the number of domains is a better proxy for the number of customers of the provider, which is valuable to include in approximating its size. 
Last, this definition also maximizes the value of our feeds as measured by their differential contribution~\cite{Pitsillidis2012}.   

Counting pairs of $\left\langle  2LD, IP \right\rangle$ mitigates the problem but is not perfect. 
In some cases it is appropriate to count $\left\langle  FQDN, IP \right\rangle$ pairs (e.g. malicious domain generation algorithms), or even $\left\langle  URL, IP \right\rangle$ pairs (e.g. child abuse content concentrated under the same domain with varying paths in the URL). 

\subsection{Data feeds}
\vspace{-5pt}
A separate decision in mapping abuse is \textit{what data feeds to use}. 
A wide range of abuse on the Internet is associated with hosting. 
Hosts are used as malware drop zones and to host phishing pages designed to steal sensitive information. 
Botnet command and control (C\&C) servers are also hosted~\cite{Bradbury2014}. 
Other types of hosting related abuse includes child pornography, SEO schemes, spam and counterfeit goods stores. 
Not all criminal activity can be observed in a way that can be attributed to the infrastructure of a specific hosting provider. 
Think of hidden services on TOR. 
Even if it can be observed, the criminal activity might not be captured in abuse data feeds, which are often produced by automated means. 
This implies that abuse feeds are always partial and of varying quality. 
This is a well known fact~\cite{Pitsillidis2012,Metcalf2013}. 
Needless to say, criminal activity that is not captured in the abuse data included in a metric, forms a blind spot of that metric. 
This suggests to include as broad a spectrum of abuse feeds as possible.

\begin{table}[b]
\vspace{-10pt}
\fontsize{6pt}{7pt}\selectfont
\caption{Data Feeds and Statistics}
\hspace{-10pt}
\begin{tabular}{lllrrrr}
\toprule
\textbf{Abuse Type} & \textbf{Feed} & \textbf{Organization}& \multicolumn{4}{c}{\textbf{Samples}}\\
\cmidrule(r){4-7}
& & & \multicolumn{2}{c}{$\left\langle  Domain, IP \right\rangle $} & \multicolumn{2}{c}{IPs} \\
\cmidrule(r){4-5} \cmidrule(r){6-7}
& & &  Total & Excl. & Total & Excl.\\
\midrule[2pt]
Malicious Hosts & \href{https://www.shadowserver.org/wiki/pmwiki.php/Services/Compromised-Website}{SHC} & Shadowserver & 3957 & 3615	& 2260 &1321 \\
Malicious Hosts & \href{https://www.shadowserver.org/wiki/pmwiki.php/Services/Sandbox-URL}{SHS}	 & Shadowserver & 7632	 & 7489	& 1100 & 816 \\
Malware & SBW	& StopBadware& 15204 & 14757 & 7702 & 6170 \\
Botnet C\&Cs & \href{https://zeustracker.abuse.ch}{ZEUS}  & Abuse.ch & 50 & 27	& 72 & 35 \\
Phishing& \href{https://www.phishtank.com/developer_info.php}{PHISH}  & Phishtank & 2278 & 1780 & 1377 \\
Phishing& \href{http://www.antiphishing.org/}{APWG}  & APWG &	3060 & 2430	& 1886 & 1101\\
Take Down Request & MLAT  & Dutch Police & 1347 & 1202 & 1433 & 1202  \\
Child Pornography& \href{https://www.meldpunt-kinderporno.nl/}{MELD}  & Meldpunt  & 725 & 584 & 417 & 242\\
\midrule[2pt]
Total & &  &  34253 & 31884 & 16247 & 11491 \\
\bottomrule
\end{tabular}
\label{tab:feeds}
\vspace{-0.4cm}
\end{table}

We collect a range of feeds and blacklists from private, public, commercial, and governmental sources. 
Table~\ref{tab:feeds} gives and overview of these data feeds.
The data spans over the entire duration of 2014 (with the exception of the SHC and SHS which span over the 2nd half of 2014). 
The majority of our feeds, do not share much information on the exact collection methodology.
We did not include some of the available spam feeds because our analysis of the data revealed these to be mostly related to compromised end users of ISPs rather than hosting companies.

In general, data quality relates mainly (but not only) to: (i) coverage (\textit{What is the overlap between the different feeds?}), (ii) purity (\textit{How much of the blacklisted domains truly host malicious content?} 
However, it is not always possible to assess the coverage or purity of a feed since mainly of its details are not well documented \cite{Pitsillidis2012}. 

\textbf{Coverage.}
Previous overlap analysis of blacklists that cover different types of abuse concludes that - although existent - there is little overlap in terms of the abuse associated with each ASN~\cite{Metcalf2013}. 
We reach similar results especially when $\left\langle  2LD, IP \right\rangle$ pairs are the unit of abuse. 

%
Clearly the feeds differ substantially in terms of the volume of reported abuse samples. 
For example, the professionally oriented \textit{SBW} feed contributed over 15,000 samples, while the non-profit \textit{ZEUS} feed three orders of magnitude less domain-IP pairs. 
In terms of the total number of IPs, \textit{SBW} reports almost two times less unique IPs than distinct $\left\langle  2LD, IP \right\rangle$ pairs whereas the \textit{ZEUS} feed reports more IPs because some Zeus config, binary, and drop zones are hosted solely on IP addresses. 
Moreover, the differences between $\left\langle  2LD, IP \right\rangle$ pairs and IPs indicate that many domains used for criminal activity are  mapped to a smaller number of IP addresses which could be the result of shared hosting services. 
Across our feeds, 93\% of all $\left\langle  2LD, IP \right\rangle$ pairs and 71\% of all IPs for \textit{all domains} were exclusive to a single feed (cf. \textit{Excl.} column in Table \ref{tab:feeds}). 
We refer to samples as \textit{exclusive} when they appear only in one feed. 

\begin{figure*}[t]
\centering
\begin{subfigure}[b]{0.4\textwidth}
\includegraphics[width=\textwidth]{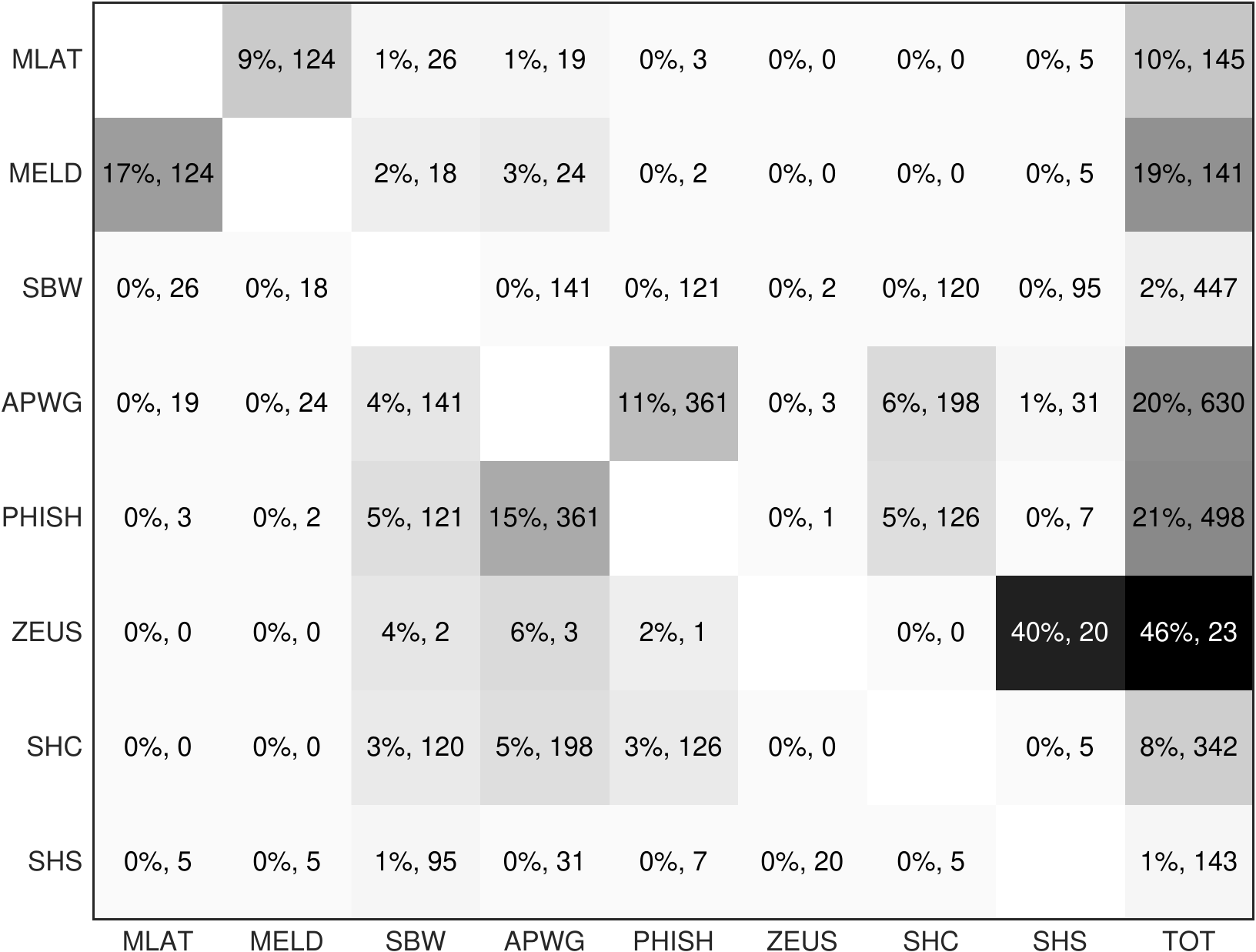}
\label{fig:overlap_dom_ip}
\end{subfigure}
\begin{subfigure}[b]{0.4\textwidth}
\includegraphics[width=\textwidth]{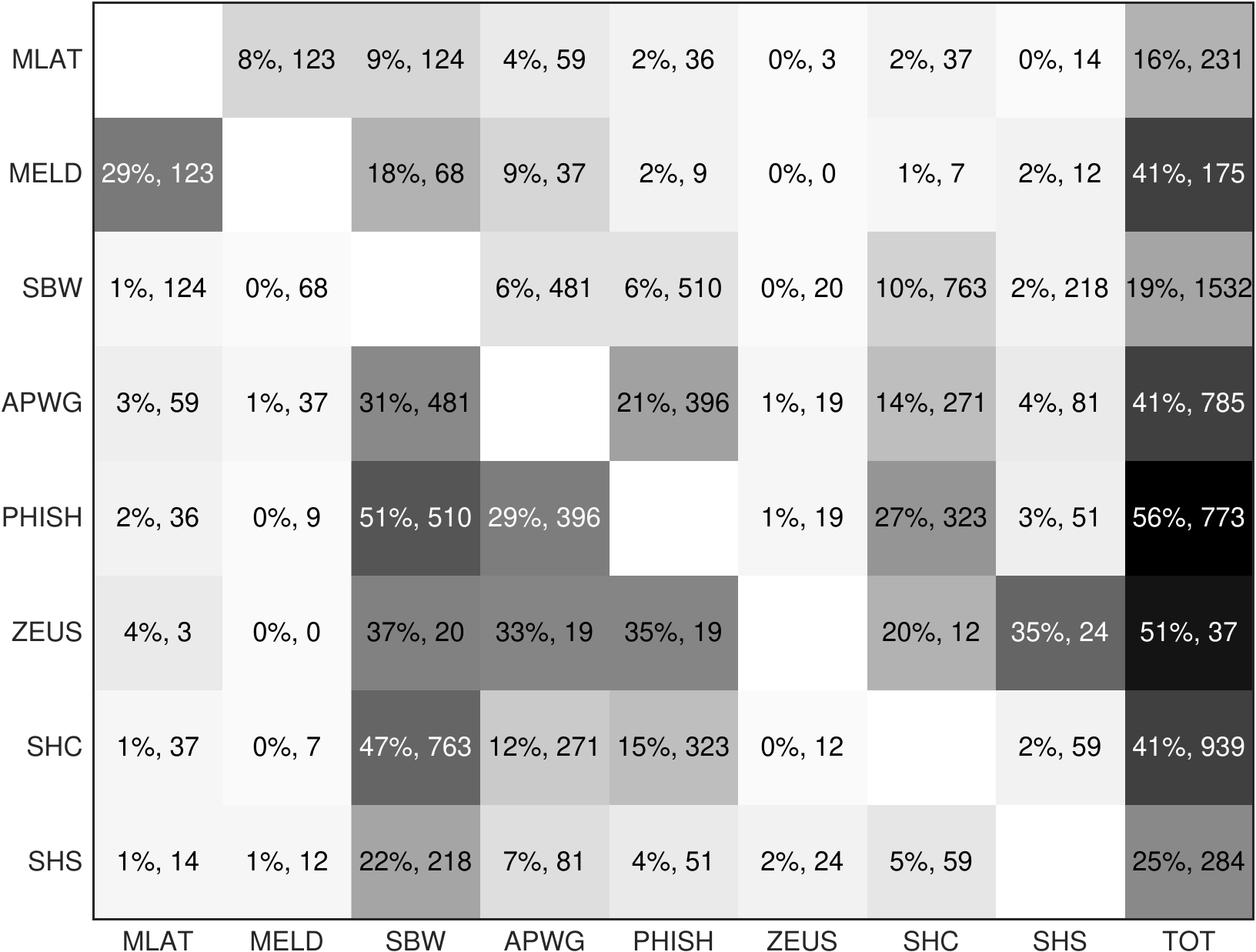}
\label{fig:overlap_ip}
\end{subfigure}
\vspace{-12pt}
\caption{Pairwise overlap of feeds with unique $\left\langle  2LD, IP \right\rangle$ (left) and IPs (right) as unit of abuse}
\label{fig:overlap}
\vspace{-10pt}
\end{figure*}

Figure ~\ref{fig:overlap} illustrates pairwise feed intersections as a matrix, with unique $\left\langle  2LD, IP \right\rangle$ (left) and unique IPs (right) as the unit for abuse respectively. 
Here darker shades of grey represent higher overlaps.
For instance, in Figure~\ref{fig:overlap} (left), the overlap between \textit{MLAT} and \textit{MELD} indicates 124 $\left\langle  2LD, IP \right\rangle$ pairs in common. 
This overlap constitutes 9\% of the MLAT feed. 
In comparison, 124  $\left\langle  2LD, IP \right\rangle$ pairs represents 17\% of the MELD feed. 
The rightmost column indicates the absolute number and the percentage of samples that the feed has in common with all other feeds combined.
The amount of overlap in the rightmost columns confirms the simultaneous use of IPs for different malicious purposes. 
This also further supports our choice of abuse unit.

Finally, the relatively small overlap among our chosen data feeds in terms of $\left\langle  2LD, IP \right\rangle$ indicates the suitability of these feeds, still, other feed characteristics need to be analyzed to further establish suitability. 

\textbf{Purity.} 
All abuse feeds contain false positives. 
The main question is which samples should be considered false positives and excluded and which should not?
We define false positives as websites maintained by legitimate users that do not serve any malicious content and are incorrectly blacklisted. 

Some domains are legitimate but point to servers that host malicious content. 
For example, we consider URL shortening services such as \texttt{goo.gl} or \texttt{bit.ly} as false positives. 
However, other legitimate websites such as free web hosting providers (e.g. Hostinger), or cloud storage services (e.g. Imagezilla.net or Dropbox) are misused by criminals and included in our analysis.

Moreover, a certain portion of abuse feeds include benign domains. 
We analyze benign domains that appear in the Alexa top 25 thousand domain list to evaluate the prevalence of false positives. 
Although we do not provide a real-time verification of blacklisted Alexa domains, we perform \textit{a posteriori} analysis to further establish the suitability of our chosen abuse feeds. 

Due to space limitation we only briefly discuss the analysis and do not include the details. 
Overall we find a low number of Alexa domains in the abuse feeds. 
Nevertheless, there are major differences among the types of Alexa domains per feed.
For example, through manual analysis of a random sample from the \textit{SHS} feed we found that approximately 30\% of this feed's Alexa domains were file sharing services most probably used to host malicious content and thus relevant to include. 
On the other hand, we also find some examples of popular websites like \texttt{msn.com}, or \texttt{microsoft.com} 
that are presumably used by compromised machines to test network connectivity. 
In the case of the \textit{PHISH} feed we find that a significant number of ranked domains that are either false positives (e.g. banks and other legitimate services), or are not appropriate for the type of analysis that we provide (e.g. URL shorteners).
The majority of the ranked domains for both \textit{MLAT} and \textit{MELD} represent file and adult content sharing services. 
As systematic false positives, or unrelated web services in our feeds do not constitute a large number we therefore have opted to include such ranked domains in our analyses.
\vspace{-7pt}
\section{Step 2 - Size Mapping}\label{sec:size_mapping}
\vspace{-5pt}
Reliable reputation metrics need to account for a commonly observed trend that larger providers also experience a larger amount of abuse. 

A common yardstick for measuring the `size' of hosting providers is the number of IP addresses routed by its corresponding AS in the BGP protocol~\cite{Stone-Gross2009,Shue2012}. 
Nevertheless not all IPs routed by an AS are used for hosting content nor are they directly in use by the AS. 
IPs may be leased and used for other purposes.
Inaccuracies in size estimation may negatively impact the reliability of a metric in that they can lead to misleading results. 
Nevertheless, due to simplicity of calculating, advertised IP space remains and attractive choice for size estimation. 

We propose (and use) two additional size estimators: (\textbf{\textit{i}}) the \textit{number of hosted 2LD}  and (\textbf{\textit{ii}}) the \textit{number of IP addresses used to host content} per hosting provider. 
To calculate these estimators we use the historical passive DNS (pDNS) data provided to us by Farsight Security.
This data records domain name resolution queries collected over the entire duration of 2014 which we use to count the number of unique 2LDs and their matching IP addresses. 
These counts are subsequently mapped to ASes routing the IP addresses and used as an estimation of the size of the provider. 
Here the quality of the resulting estimates is highly dependent on the coverage of the pDNS data. 
As long as the pDNS data has a reasonable coverage of all registered 2LDs, it can be used to produce reasonable size estimates. 
We have crosschecked the number of unique 2LDs observed in the pDNS data with the number of 2LDs of gTLDs and ccTLDs present in zone files of new gTLDs that we obtained under agreement from ICANN~\footnote{\url{http://newgtlds.icann.org}} in addition to ccTLD sizes reported by APWG~\footnote{\url{http://docs.apwg.org}} at the end of 2014.
Extrapolating from these results we have concluded pDNS to be a reasonably reliable source to estimate hosting provider size.

There are potentially other conceivable estimators for hosting provider size such as the number of customers. 
Nevertheless, the scarcity of data to base such estimates on is a largely limiting factor in this respect.
\section{Step 3 - Normalization of Abuse}\label{sec:normalization_abuse}
\vspace{-5pt}
Given the output of abuse mapping and size mapping, the next step in the metric production process is to normalize abuse by a size estimate.
This leads to $S \times N$ normalized abuse mappings where $S$ is the number of size maps produced earlier and $N$ the total number of abuse maps 
corresponding to analyzed blacklists.
A key question here relates to \textit{interpretations that can already be made from normalized abuse data}.

All size estimates have their advantages and disadvantages which have to be viewed as trade-offs. 
The most commonly used size estimator - routed IPs - is the easiest to calculate, but it suffers from systematically favoring large providers, since not all routed IPs are used for hosting.
Using the portion of the routed IP space that is used for hosting as the size estimator mitigates the problem, however, this is much more difficult to calculate. 
This estimate is also not free of systematic bias, because it favors hosting providers that have a disproportionately large amount of shared hosting.  
We can use the number of hosted 2LDs as the estimator, which would treat shared hosting fairly but would still underestimate the size of subdomain resellers and free-hosting services.
The trend here is clear; normalized abuse has its blind spots, and needs to be taken into account especially for interpreting results at this stage.  

It is important to note that some size estimates are more volatile than others due to the dynamic nature of the underlying processes. 
For example, the number of FQDNs hosted by a provider may change at a much faster rate than the number of 2LDs if an estimator based on FQDNs is used.

Normalized abuse, is already an indicator of security performance by itself. 
Note however, that normalized abuse is abuse type specific. 
For example, one can analyze normalized abuse based on the occurrence of malware on hosting providers and draw conclusions; however, this only provides a partial picture of the performance of hosting providers. 
Some providers might be much less strict about allowing malware spread from their servers than for example the hosting of child pornography~\cite{Bradbury2014}. 
In our case, we use all size estimators outlined in the previous section without committing to a specific one or considering one superior to others. 
The expectation is that the combination of these can overcome the deficiencies of each.
This matter is further explored in Section~\ref{sec:sensitivity}. 

Finally, note that when talking about metrics based on the up-time of abuse, size corrections are not appropriate. 
In such cases it is common to use mean or median up-times instead of normalized abuse.
\section{Step 4 - Rating of Abuse}\label{sec:rank_abuse}
\vspace{-5pt}
Given the normalized abuse maps, the next step in the process, calculates rankings over all maps to produce rankings. 
Rankings are one way of unifying the scales on which normalized abuse is measured and allows cross comparisons over categories of abuse. 
For example, comparing the security performance of a hosting provider in terms of how well it manages to mitigate malware with its performance in terms of how well it mitigates phishing is not meaningful when based on normalized abuse. 
However, the comparison is meaningful over rankings.

Given the normalized abuse maps, our method for ranking hosting providers is as follows:
We rank normalized abuse from high to low.  
This results in  $3 \times N$ \textit{rankings}. 
The individual rankings may range between zero and $R$, the total number hosting providers. 
The worst rank, $R$, is assigned to the AS with the highest normalized abuse, $R-1$ to the second worst and so forth. 
ASes with equal normalized abuse are assigned equal ranks. 
If  a normalized abuse map only contains data on for example 20 providers the ranking will range between $\left[R-20, R\right]$ with all providers for which no abuse was detected receiving the low rank of $R-20$. 

An important consideration in producing rankings is \textit{information loss}.
To illustrate this consider two hosting providers $HP_1$ and $HP_2$ that have a normalized abuse of $0.1$ and $0.3$ and have been assigned the ranks of 10 (worst performer) and 5 (5th worst) respectively. 
Our ranking is not \textit{distance preserving} since the difference between the hosting provider ranks ($10 - 5$) does not entail the same information as that of the normalized abuse ($0.3 - 0.1$ ). 
That is, one unit of change in ranking could mean any number of changes in the unit of normalized abuse. 
As a result these distances cannot be interpreted in the same way. 
In ranking hosting providers, some information about the magnitude of the differences is unavoidably lost.
\vspace{-10pt}
\section{Step 5 - Aggregation of Rates}\label{sec:combine_ranks}
\vspace{-5pt}
We now aggregate our rankings into one overall ranking that assigns scores in the range $\left[0, 1\right]$, where score 1 indicates the worst performer. 
The aggregation procedure considers every ranking as a voting preference over $R$ candidates in an imaginary election. 
The election winner is effectively decided using a \textit{Borda Count} vote aggregation method that basically counts how many times a certain candidate appeared in the 1st place, 2nd place, 3rd place (and so forth) in every ranking and decides the outcome based on all rankings. 

An alternative approach could perform factor analysis and take into account the most contributing feeds when interpreting metric scores (cf. also Section \ref{sec:sensitivity}). 
%
We find, however, voting systems to be a useful analogy when thinking about aggregation. 
A useful aggregation method must have certain desirable properties, such as being intuitive.
For example, if a particular hosting provider is the worst ranked performer in all categories of abuse, the reputation metric should reflect that by assigning the worst metric score to that provider and not to others. 
Certain methods of aggregation will not guarantee such properties and are therefore undesirable. 
We refer the reader to literature on different voting aggregation methods~\footnote{See \textit{e.g.} \url{http://lorrie.cranor.org/pubs/diss/node4.html}}  for a better understanding of the properties of such methods and their limitations.
\vspace{-5pt}
\section{Step 6 - Metric Interpretation}\label{sec:interpretation}
Reputation scores need to be interpreted to guide policy and reduce information asymmetry around the security performance of hosting providers. 
However, correct interpretation of a metric without detailed knowledge of the various blind spots and biases of the process is difficult. 
Additionally the heterogeneity in the hosting provider landscape directly influences what conclusions can be drawn from the scores. 

\begin{figure}[tb]
\centering
\includegraphics[width=\linewidth]{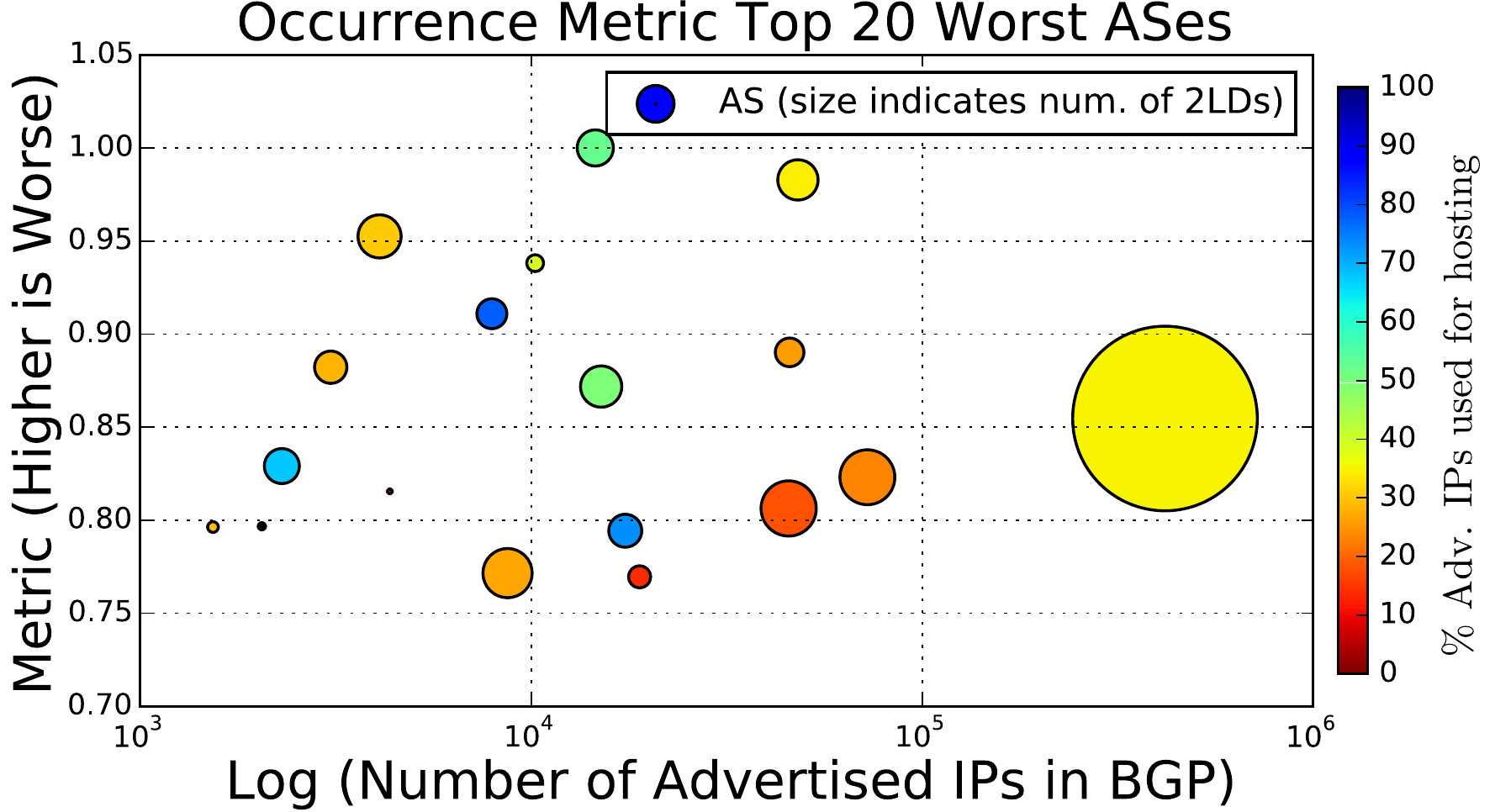}
\caption{The 20 worst Dutch providers for abuse rate. }
\vspace{-15pt}
\label{fig:occurrence_metric}
\end{figure}

To illustrate the challenges of interpretation we briefly present some of our results here.
Figure~\ref{fig:occurrence_metric} plots the rankings of the 20 worst performers based on the occurrence of abuse. 
The plot demonstrates a large variance between the performance of providers that have comparable sizes.
The results clearly indicate significant differences in how hosting providers deal with abuse. 
Here, the safest comparisons are among providers that have the most similar properties. 
As an example consider the two hosting providers colored in bright green, first the provider with the highest metric score and second the provider located approximately at $(x \backsimeq 10^4, y \backsimeq 0.85 )$.
These are very similar in all aspects and therefore it can be safely concluded that the provider with the lower score is performing significantly better than the worst performer due to its security policies and practices. 
To consider the worst provider as negligent or criminally engaged simply because it has the worst score is however a wrong conclusion to draw here. 

\begin{figure}[tb]
\centering
\includegraphics[width=\linewidth]{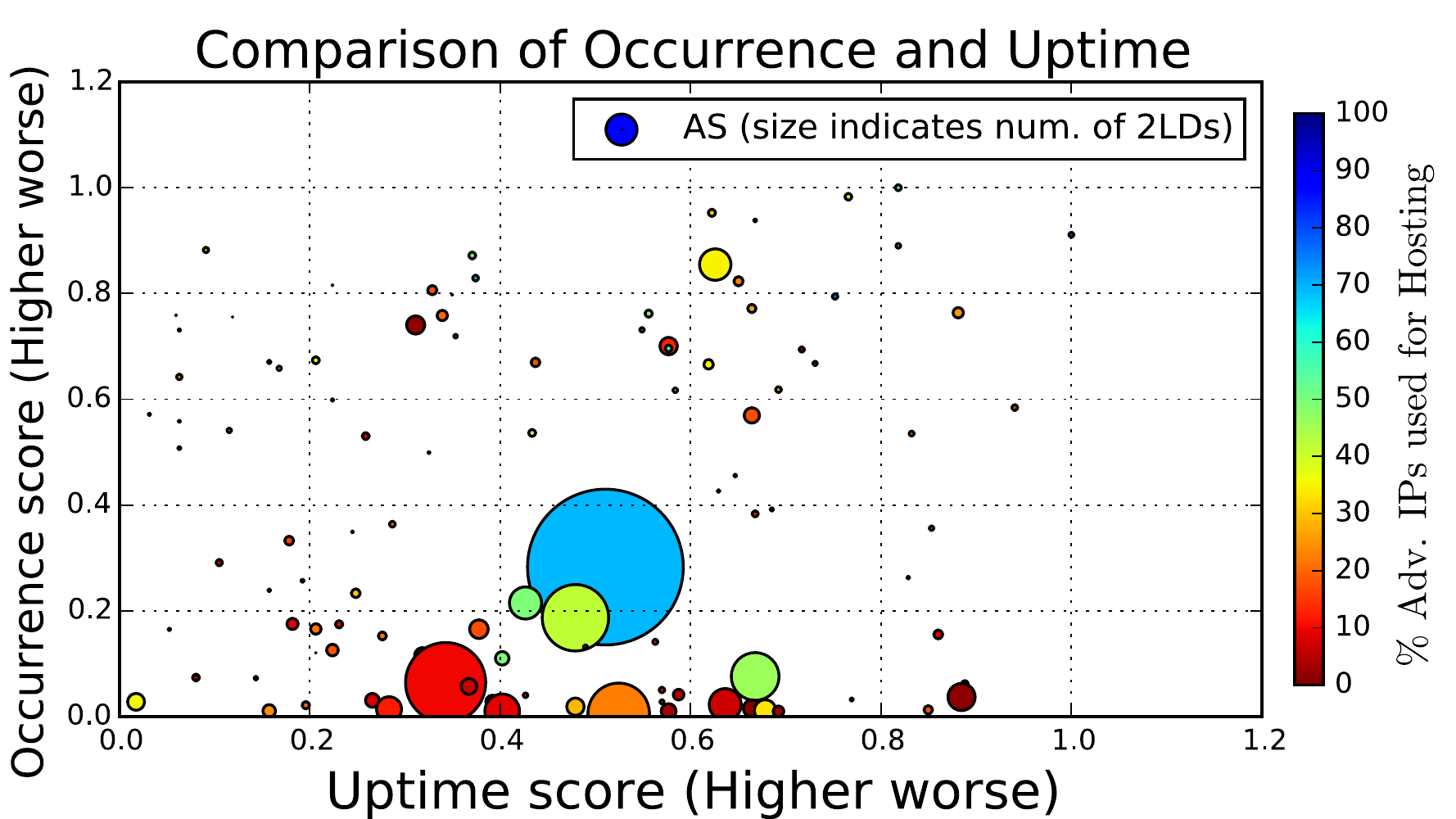}
\caption{Comparison of occurrence and uptime metric.}
\vspace{-17pt}
\label{fig:comparison_metric}
\end{figure}

Figure~\ref{fig:comparison_metric} compares the occurrence metric and the uptime metric of hosting providers. 
A cautionary note here is that our uptime metrics are based on only 2 data feeds from which up-times could be calculated. 
The weak relationship (Spearman's  $\rho=0.38$, Pearson $r=0.36$) between occurrence and up-time is expected as each captures a different aspect of hosting provider characteristics that relate to abuse (see Section~\ref{sec:overview}). 
Clearly some providers experience large amounts of abuse while managing to quickly block the abuse (upper left region of the plot). 
Others, perform consistently bad in the sense that they experience large abuse occurrence and are also slow to block it (upper right region of the plot).
Nevertheless, we believe that the amount of occurring abuse and the response of a hosting provider to abuse are important aspects that need to be both measured separately to provide a thorough picture of security performance. 
Only now can we draw the conclusion that the worst performing hosting provider in terms of occurrence is probably negligent because it is also among the worst performers in terms of uptimes (see point with $(x \backsimeq 0.8, y = 1 )$ coordinates)

Finally when interpreting the results, one should also take the hosting provider business model into account. 
Hosting providers with a large portion of shared hosting customers have a larger role to play in cleaning up content than ones with dedicated hosting clients. 
It might very well be the case that the worst performer in terms of both occurrence of abuse and response to abuse provides solely unmanaged hosting to its customers and therefore not in the same position as its peers that provide mainly shared hosting. 
In this case the observed performance could simply be indicating the security of the hosting customers rather than that of the hosting provider itself. 

\vspace{-10pt}
\section{Sensitivity Analysis}\label{sec:sensitivity}
\vspace{-10pt}
To better understand the impact of key design decisions, we undertake a brief sensitivity analysis of alternative specifications for \textit{(i)} unit of abuse, \textit{(ii)} abuse normalization, and \textit{(iii)} metric aggregation strategies.
We explore the robustness of the results by producing rankings based on alternate methodological options for each of these decisions. We compare them to the ranking of the pragmatic approach we presented (called as benchmark ranking) by calculating the \textit{Pearson correlation coefficient} among the top $n=100$ worst performing Dutch hosting providers.

\textbf{Unit of abuse.} 
We used unique $\left\langle  2LD,IP \right\rangle$ pairs as the unit of abuse. We calculated an alternate ranking based on \textit{unique IP counts}, the standard approach in the literature. 
The Pearson's $r$ among both rankings is $0,952$. Perhaps more tangible: 16 ASes are in the top-20 of worst performers of both rankings. In other words, the metric is not very sensitive to either specification.

\textbf{Abuse normalization.}
We calculated three alternate scores using the three size estimates for hosting providers: \textit{(i)} the advertised IP space \textit{(ii)} the advertised IP space that is used for hosting, and \textit{(iii)} the number hosted 2LDs. 
The benchmark ranking used all three of them. 
The Pearson's $r$ for the alternative specifications to the benchmark ranking are $0,896$, $0,909$, and $0,6438$, respectively. 
The results reveal a strong correlation between the IP space-based size estimators and the benchmark ranking, and a less strong correlation with the estimates based on 2LDs. 
Out of the top 20 worst performers in the benchmark ranking, only 7 ASNs were present in all alternate top 20 rankings. When comparing pairwise: the reference ranking shares 13 ASNs in common with those based on advertised IP addresses and hosted 2LDs, and 15 ASNs with the ranking based on IP addresses used for webhosting.
In other words, using the number of hosted domains vs. estimates based on IP address space give significantly different results. 
By including all three size estimations, our metric specification mitigates that impact, while retaining the advantages of including domain name counts in abuse counting and size estimation.

\textbf{Metric aggregation.} We also compared our benchmark ranking with a ranking in which we assigned weights to each data source according to their comprehensiveness, i.e., the relative volume of each feed in terms of distinct number of exclusive $\left\langle  2LD,IP \right\rangle$  pairs in the dataset (cf. Table \ref{tab:feeds}). 
In summary, out of top 20 abused ASNs in the benchmark, 14 ASNs showed up in the weighted rankings and the Pearson's r of the benchmark ranking and weighted ranking is equal to $0,791$.
\vspace{-5pt}
\section{Related Work}
\vspace{-5pt}
Numerous studies have pointed to concentrations of abuse in certain networks, typically in the context of a specific criminal business model, e.g., spam~\cite{Eeten2010}, phishing~\cite{Moore2007} or malware~\cite{Kuhrer2014}. 
Effects and policy implications of intervention at classes of intermediaries have also been studied in~\cite{Eeten2010,Hao2013}.
The quality of abuse data has also been extensively covered~\cite{Pitsillidis2012,Metcalf2013,Kuhrer2014}. 
\cite{Canali2013b} examines the role of hosting providers in detecting abuse and reacting to user complaints for shared hosting providers. 
It paints a general picture that underlines the need for hosting reputation metrics. 
None of these studies try to develop reputation metrics from abuse data, however.

Closest to our work are~\cite{Stone-Gross2009,Shue2012,Gros}.  \cite{Wagner2013} produces a weighted metric score for ASes. 
\cite{Stone-Gross2009} includes only up-time data and focuses mainly on identifying the worst actors.
We expand on this work by systematically addressing challenges not discussed there. 
These studies typically count IP addresses as the unit of abuse and use advertised IP address space as a normalization factor.  
 
Industry attention to hosting has been along the lines of the Host Exploit Index (HE index)~\cite{hostexploit}. 
While valuable, their methodologies are not fully transparent and the parts that are, suffer from similar limitations as the academic work discussed above.
\vspace{-10pt}
\section{Conclusions}
\vspace{-5pt}
This paper has systematically worked through the many challenges of developing security reputation metrics for hosting providers. 
All conceivable metrics will suffer from various limitations, that much is clear. 
This is not to say that they are not useful. 
We presented our approach to various stakeholders in the Netherlands, including hosting providers. 
The main response was that the metrics were a valid starting point for evaluating hosting security, incentivizing self-regulation and, ultimately, identifying actors for enforcement activities. 

The way forward is to improve the methodology. 
First, we aim to include additional abuse data feeds and more uptime data. 
We also are improving the identification of hosting providers by using WHOIS data on IP address ownership, rather than AS-level routing data. 
We are working on techniques to differentiate various hosting provider services, so that abuse rates can take these into account. 
A further step is to use different aggregation techniques, such as factor analysis. 
We also aim to further investigate incentives under certain conditions where security metrics can be gamed by providers. 
Last, but not least, we will undertake more in-depth sensitivity analysis of how the various methodological decisions impact the metric. 

It is safe to say that a lot of current claims about hosting providers are based on anecdotal evidence or methods that are not adequately understood. 
This paper contributes to remediating this shortfall.

\subsubsection*{Acknowledgements}
\fontsize{8pt}{9pt}\selectfont
This work was funded by NWO under Pr. Nr. CYBSEC.12.003 / 628.001.003,  and SIDN (www.sidn.nl). We would like to thank Paul Vixie and Eric Ziegast for great support, NCSC, the Dutch police, Farsight Security, StopBadware, APWG, Meldpunt, Shadow Server, Abuse.ch, and PhishTank for providing access to their data.

\printbibliography
\end{document}\grid